\documentclass[11pt]{article} 
\usepackage{natbib}
\usepackage{rldmsubmit,palatino}
\usepackage{graphicx}
\usepackage{amsmath}
\usepackage{amsfonts} 
\usepackage{subcaption}

\title{Subject Matter Expertise vs Professional Management in Collective Sequential Decision Making}

\author{
David Shoresh \\
The Edmond and Lily Safra Center for Brain Sciences\\
Hebrew University\\
Jerusalem \\
\texttt{david.shoresh@mail.huji.ac.il} \\
\And
Yonatan Loewenstein \\
The Edmond and Lily Safra Center for Brain Sciences \\
Departments of Neurobiology and Cognitive Sciences
and the Federmann Center for the Study of Rationality \\
Hebrew University \\
Jerusalem \\
}

\begin{document}

\maketitle

\begin{abstract}
Your company's CEO is retiring. You search for a successor. 
You can promote an employee within the company, who is familiar with the company’s operations, or recruit an external professional manager from a foreign industry. 
Who should you prefer?
It has not been clear how to address this question, known as the subject matter expertise vs. professional manager debate, quantitatively and objectively.

To address it, we note that the company's success depends on the manager making long sequences of interdependent decisions, relying on the often-opposing recommendations of diverse team members. 
To model this task in a controlled environment, we utilize chess, a complex, sequential game with interdependent decisions. 
Chess allows for quantitative analysis of performance and expertise because the states, actions and game outcomes are well-defined.
The availability of chess engines that differ in style and expertise, allowed scalable experimentation.
We considered a team of (computer) chess players. At each turn, team members recommend a move and the manager chooses one recommendation.
We compared the performance of the different types of managers.
As a subject matter expert, we used another (computer) chess player that assesses the recommendations of the team members and chooses from them based on its own chess expertise. 
We examined the performance of such managers at different strength levels.
To model a professional manager, we used Reinforcement Learning (RL) to train a network that identifies the board positions in which different team members have relative advantage, without any pretraining in chess. 
We further examined this network to see if any chess knowledge is acquired implicitly.
We found that subject matter expertise beyond a minimal threshold does not significantly contribute to team synergy. 
Moreover, performance of a RL-trained professional manager significantly exceeds that of even the best expert managers, while acquiring only limited understanding of chess.
\end{abstract}


\acknowledgements{This work was supported by the Gatsby Charitable Foundation (YL).
Y.L. is the incumbent of the David and Inez Myers Chair in Neural
Computation.}

\startmain 

\section{Introduction}
The management literature is divided on the dilemma of which is more important for effective management: subject matter expertise or professional management skills \citep{Phillips}. 
The research in this field has suffered from various problems. For example, those using quantitative methods have had to rely on indirect measurements of expertise and professional management skill \citep{Graf}.
Furthermore, studies have emerged that support both positions. 
For example, one study found that doctors make better hospital managers than non-medically trained administrators \citep{Doctors}. Similarly, \citep{Adams} reported that deep functional expertise by CEOs positively affect financial performance. On the other hand, studies of boards with diverse expert members have shown the importance of developing a "transactive memory system" in which there is a "conceptualization of ‘who knows what’"\citep{Brandon}. This indicates the importance of being able to delineate relative advantages in diverse teams based on experience.
Part of the problem is that management is a multifaceted phenomenon, making it hard to disaggregate causes.
In this work we attempt to study this problem in a controlled, simulated environment, with direct empirical measurement of expertise and with a clearer distinction between manager types. We do this using chess.

\section{Experimental setup}
\subsection{Team players}
We deploy two types of chess models to play as a team. 
One is Leela-Chess-Zero ("Leela"), an open-source version of Alpha-Zero \citep{Silver}, which is a pure self-play reinforcement learning algorithm. 
The other is Maia-Chess ("Maia"), an open-source family of models trained from human games \citep{McIlroy}. 
Maia and Leela play as a team against a popular chess engine known as Stockfish, which has a heuristics-based evaluation function, parameterized by a shallow neural network. 
There are multiple Maia, Leela and Stockfish versions available online. 
In all our experiments we used a Maia network ranked 1900 ELO (comparable to a strong pre-master human ranking in chess) when deployed at search depth 1, and a comparable level Leela network. The version of Stockfish that we used is slightly stronger.

\subsection{Team protocol}
At each board position \emph{s} arrived at during a game, if Maia and Leela agree on the best move, that move is played. 
If they disagree, a manager chooses from their recommendations. 
If the manager is indifferent, the move is chosen randomly from the two recommendations.
Moves are chosen in this way from the opening position (chosen from a library of open positions) until the end of the game.

To formalize a "team", we define a team policy, $\hat{\pi}$, that consists of team member policies ($\pi_1, \pi_2$), and a selection mechanism between them at each state which we denote the "manager".
The manager can decide based on the state and move recommendations ($\text{Manager}(s, a_{1}, a_{2})\rightarrow k\epsilon\left[1,2\right]$), or it can decide based only on the state ($\text{Manager}(s)\rightarrow k\epsilon\left[1,2\right]$). In the latter case, the manager chooses whose recommendation to follow without knowing what these recommendations are. 


Note that it is trivial to construct a manager that will perform as well as the better of the two team members -- simply adopt all its recommendations. The challenge is to achieve "synergy" -- manage the team such that its performance exceeds that of either team member alone (measured empirically when each plays separately against the same adversary).



For evaluation, we used a set of chess opening positions, and ran the team against the adversary from both black and white perspectives for each position. To rate performance, we use the “Wins-Draws-Losses” score (WDL) defined as:

\[
  \text{WDL}=\frac{\text{\#wins}+\frac{1}{2}\cdot \text{\#draws}}{\text{\#wins}+\text{\#draws}+\text{\#losses}}
\]



\section{Subject matter expert}

To model a subject matter expert (SME), we use another chess engine as the manager. 
This "expert" has access to the recommended moves of Maia and Leela, and it scores out each of them using its own evaluation function. The chosen action is the one associated with the highest score.


Note that in a non-sequential decision task, the strongest expert is, by definition, the best possible manager.
In a sequential decision task this is less clear.
This is because the assumption when evaluating the two actions is that after the chosen move, the subject matter expert will continue alone. This is not what happens in the team, because throughout the game, the manager makes decisions only when the team members disagree, and the only actions that it considers are the two actions proposed by the team members.

To simulate the expert manager, we used a set of chess programs, some of which were stronger than the team members, but also some that were weaker than them. These managers included Leela and Maia networks of variable strength, and Stockfish engines set to various search depths. To evaluate the expertise of the managers, we tested their performance when playing alone (not as a manager) against the same adversary as the team. Evaluations were over 10000 games from randomized opening positions.

\begin{figure}
  \centering
  \includegraphics[width=0.7\linewidth]{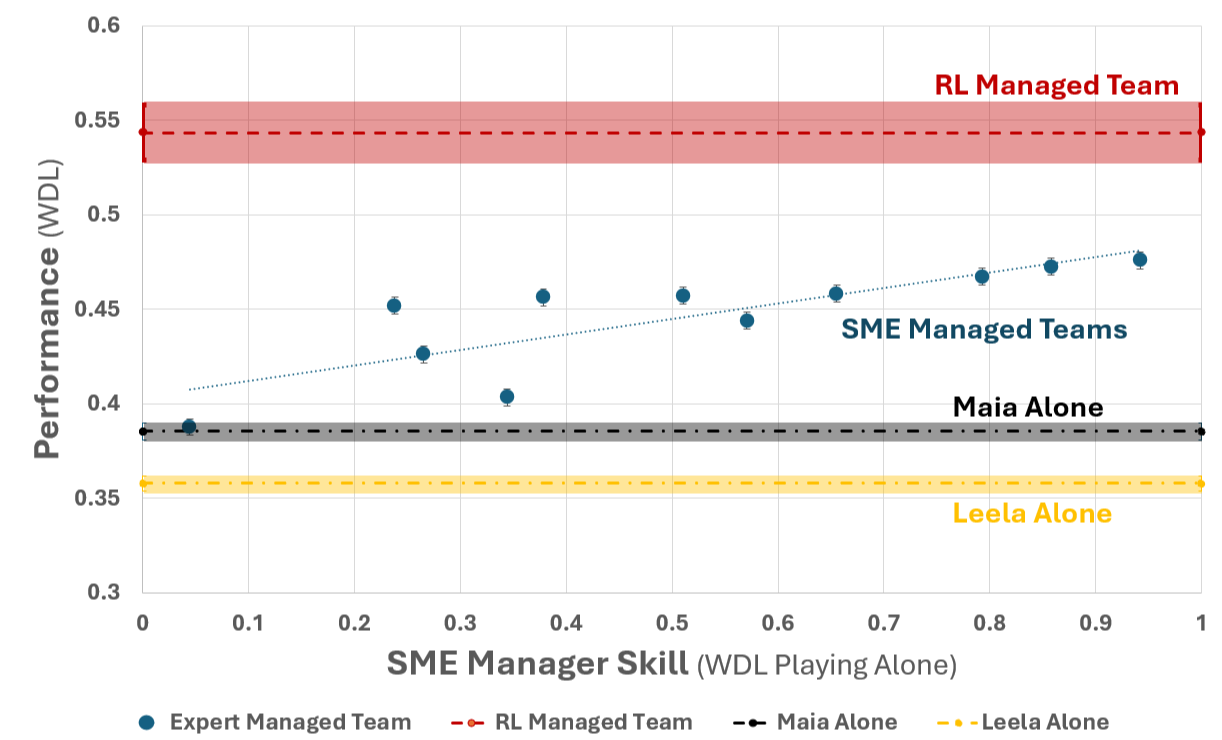}
  \caption{Team performances. The blue points show performance of teams led by subject matter experts (SME), of varying strength. The trend line has slope 0.0823 with p-value of 0.005. Confidence interval for the slope at 95\% confidence is between 0.032 and 0.132. Teams outperform the strongest team member playing alone (black dashed line). However, these "expert" managers do not reach performance level of the RL-trained manager (red dashed line). All error bars and error shades represent standard error of the mean.}
  \label{ExpertsResults}
\end{figure}

Figure \ref{ExpertsResults} depicts the performance of the team as a function of the expertise of the manager (blue dots). For comparison, the performance levels of Leela and Maia alone are denoted by the yellow and black dashed lines, respectively. One interesting observation is that the four left-most dots depict teams that were managed by an expert that was weaker than the team members, yet even these weak managers produced team synergy, (except for the weakest one). 
This can be explained by ensemble dynamics, whereby even weak decision-makers can improve overall performance.
It is also interesting that even a substantial increase in the manager's expertise had only a small effect on the team's performance. For example, the WDL score of the team led by the strongest expert manager (WDL 0.942 when playing alone) was 0.476, only 5.73\% higher than 0.452, the WDL score of the team led by the second-weakest manager (WDL of only 0.238 when playing alone).
This may be due to the "curse of knowledge", \citep{Curse},
which refers to agents with superior knowledge that fail to account for the behaviors of agents with inferior knowledge, and therefore make suboptimal cooperative decisions.

\begin{figure}
  \begin{subfigure}{0.5\textwidth}
    \includegraphics[width=0.9\linewidth]{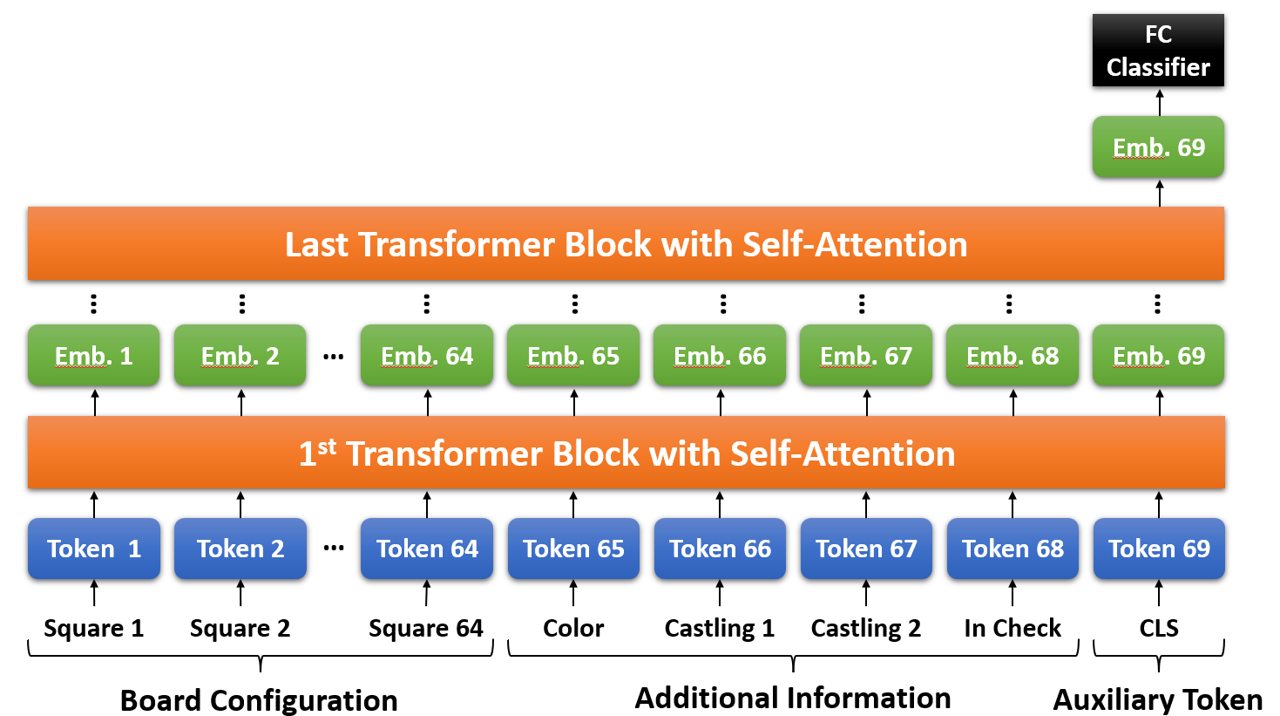}
    \caption{Transformer model}
    \label{architecture}
  \end{subfigure}
  \begin{subfigure}{0.5\textwidth}
    \includegraphics[width=0.9\linewidth]{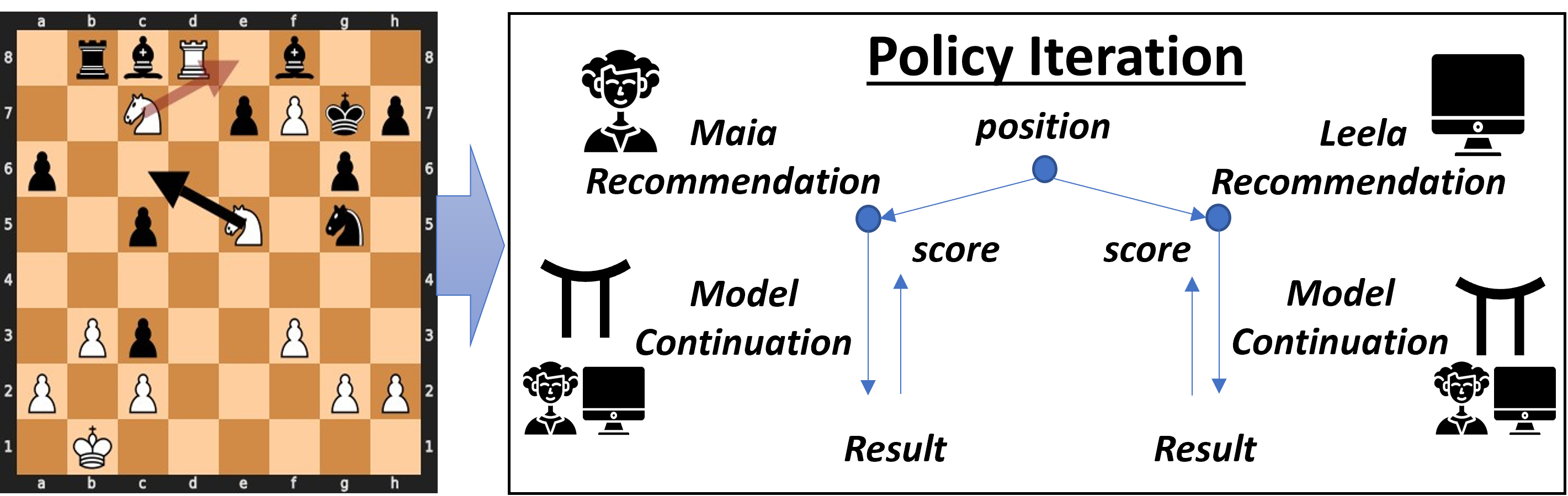}
    \caption{Process for obtaining estimated \emph{Q} values}
    \label{PolicyIteration}
  \end{subfigure}
  \caption{Training the RL manager - left: network architecture, right: training algorithm.}
\end{figure}

\section{RL-trained professional manager}

As a model of professional management, we considered a manager that is naive about the game of chess, but can identify the relative advantages of the different team members.
For this, we used RL to learn relative advantages of the team players, using game outcomes as reward. Naturally, professional management is comprised of a complex set of skills, and this only captures one of its aspects, albeit an important one.

The RL manager was parameterized by a transformer (illustration in figure \ref{architecture}). 
The model receives a representation of the board using 64 tokens, one for each square. 
The tokens denote the pieces on each square, or if it is empty, plus positional encoding. 
Additional tokens signify the color being played, castling rights and whether there is check. 
A CLS token in the last layer is fed to a fully-connected classifier network which outputs a binary indicator for Maia or Leela. 
The transformer network was much smaller than the convolutional networks used by the Maia or Leela chess engines.

To train this manager, we used policy iteration. 
At each iteration, we played a set of games, recording all disagreements between Maia and Leela. 
From each disagreement, Maia and Leela recommended moves were rolled out separately, with the rest of the game controlled by the current RL manager. 
Outcomes of these rollouts served as reward (see figure \ref{PolicyIteration}). Note that the RL manager receives only the state as input, and does not have access to the recommended moves. 
This puts it at a disadvantage relative to the subject matter expert.
Moreover, it does not have any explicit training in chess. 

The RL manager was evaluated over 1000 games from a fresh set of opening positions and achieved WDL=0.54, significantly better than the strongest subject matter expert team ($Z=2.95$). This is indicated by the dashed red line in figure \ref{ExpertsResults}.
The result indicates that in this management problem, identifying the relative advantages of the team players is more important than subject matter expertise.

\section{Is domain knowledge necessary to identify relative advantage?}

Despite the fact that the RL manager was not exposed to chess rules or strategy, it might have still learned some of these implicitly (just as a human professional manager may gain some subject matter expertise while on the job). 
We explored whether the RL manager used non-trivial understanding of the domain (chess) to make its decisions by examining the attention scores in the transformer network. 
Attention scores are used in transformer models to weight input tokens (in our case squares on the chess board) during processing, which is sometimes leveraged for explainability \citep{Vaswani, Wiegreffe}.

\begin{figure}
  \centering
    \includegraphics[width=0.9\linewidth]{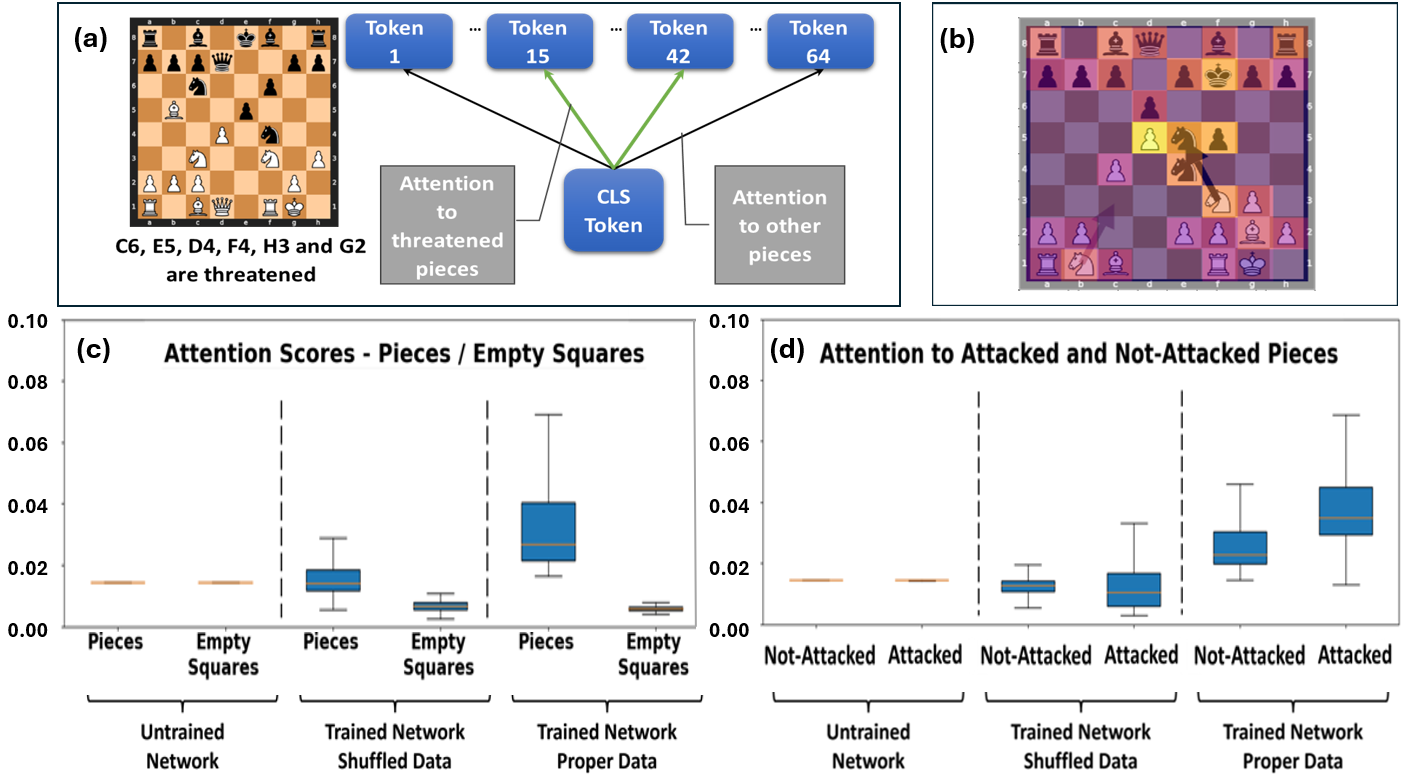}
    \caption{Examination of RL-manager internal network. (a) Illustration of  attentions extracted from the model; 
    (b) Example attention heatmap overlay (lighter colors denote higher attention);
    (c) Comparison of attentions to pieces vs empty squares;
    (d) Comparison of attentions to attacked pieces vs not-attacked.
    Whiskers in box plots represent farthest data point within 1.5x the inter-quartile range from the box.
    }
  \label{DomainKnowledge}
\end{figure}

To see whether the RL manager learned something about chess, we checked if the network gives more attention to pieces than to empty squares. 
Over 8,000 chess positions were fed to the network.
Attention scores of the CLS token towards tokens representing squares on the board were recorded (averaged over layers and attention heads).
For each position, the mean attention to pieces and to empty squares were taken. 
For control, we compared to the same attentions given by an untrained network.
Since pieces are correlated with regions of the board, we used an additional control - the trained network attentions over shuffled chess positions. Results are shown in figure \ref{DomainKnowledge}(c).
In the untrained network, attentions are near uniform.
With the shuffled data, the trained network gives more attention on average to pieces than empty squares. 
With the correctly ordered data, the distributions of attention to pieces and to empty squares hardly overlap. 

For a more advanced test, we checked if the network gives more attention to pieces that are attacked, than those that aren't, even though the network was never shown how pieces move in the game. 
Once again, we controlled with attention in an untrained network, and attention in the trained network over shuffled positions. Results are shown in figure \ref{DomainKnowledge}(d).
As before, the untrained network gives near-uniform attention to all squares. 
With the shuffled data, there is no preference for attacked pieces over non-attacked. 
But over the ordered data the trained network clearly gives more attention to attacked pieces than not. 
We measure effect size using non-parametric $A_w$ which is a robust test for group differences in numeric variables \citep{Li}. 
It gives a high effect size of 0.81 (1 is highest).

Our results indicate that after training, the model locates the pieces and observes their interrelationships in ways that correspond with the rules of chess. 
This indicates that while learning Maia and Leela relative advantages, the manager obtained some minimal domain knowledge.

Incidentally, we also analyzed the board positions in which Maia and Leela were given relative advantages, but were unable to identify simple features that were predictive of these choices.

\section{Conclusions}
Collective sequential decision making is a key aspect of management. 
Yet it is not clear how to resolve conflicting policy recommendations in a team. 
Theoretically, there is serious potential for failure, since switching between policies could prevent a coherent plan.
We find empirically that managers can succeed in producing a synergistic team policy, and that they don't need deep subject matter expertise to do so.
Our novel methodology may be a simplification of the management setting, but it provides a unique opportunity for empirical investigation of this important question.

\bibliographystyle{abbrv}
\bibliography{output}
\end{document}